# Effects of Functionalization on Thermal Properties of Single-Wall and Multi-Wall Carbon Nanotube – Polymer Nanocomposites

Richard Gulotty[1], Micaela Castellino[2], Pravin Jagdale[2], Alberto Tagliaferro[2] and Alexander A. Balandin[1*]

[1] Nano-Device Laboratory, Materials Science and Engineering Program, Bourns College of Engineering, University of California – Riverside, Riverside, California 92521 USA

[2] Applied Science and Technology Department, Polytechnic of Turin, Turin, 10129 Italy

### Abstract

Carboxylic functionalization (-COOH groups) of carbon nanotubes is known to improve their dispersion properties and increase the electrical conductivity of carbon-nanotube – polymer nanocomposites. We have studied experimentally the effects of this type of functionalization on the thermal conductivity of the nanocomposites. It was found that while even small quantities of carbon nanotubes (~1 wt%) can increase the electrical conductivity, a larger loading fraction (~3 wt%) is required to enhance the thermal conductivity of nanocomposites. Functionalized multi-wall carbon nanotubes performed the best as filler material leading to a simultaneous improvement of the electrical and thermal properties of the composites. Functionalization of the single-wall carbon nanotubes reduced the thermal conductivity enhancement. The observed trends were explained by the fact that while surface functionalization increases the coupling between carbon nanotube and polymer matrix it also leads to formation of defects, which impede the acoustic phonon transport in the single wall carbon nanotubes. The obtained results are important for applications of carbon nanotubes and graphene flakes as fillers for improving thermal, electrical and mechanical properties of composites.

---

* E-mail (AAB): balandin@ee.ucr.edu

Carbon nanotubes (CNTs) and graphene attracted attention as fillers in various types of nanocomposite (NC) materials.[1-4] One can expect that the high intrinsic electrical conductivity, $\sigma$, thermal conductivity, $K$, and mechanical strength of CNTs would lead to a substantial increase in the average electrical and thermal conductivities of the resulting composites at small loading fractions of CNTs. It was demonstrated that addition of CNTs to polymers can increase the electrical conductivity by many orders of magnitude from $10^{-10} - 10^{-5}$ to $10^1 - 10^4$ S/cm.[5]  The electrical percolation thresholds, $f_T$, for single-wall carbon nanotube (SW-CNT) composites was observed at very low CNT loading fraction, $f\sim0.1$ vol. %, as compared to 20–30 vol. % for composites with spherical metallic fillers.[6-7] The effects of CNTs on the heat conduction property of composites is characterized by the thermal conductivity enhancement factor defined as

$$\eta = (K - K_m) / K_m \qquad (1)$$

where $K$ and $K_m$ are the thermal conductivities of the composite and the matrix material respectively.

The reports on the thermal conductivity of composites after addition of CNT were less impressive and rather inconsistent. There have been studies that reported the thermal conductivity enhancement factors in the range ~50-250% at $f\sim7$ vol. % of the CNT loading.[8-10] In other cases, $K$ was not improved[11] or even decreased with addition of SW-CNTs.[11] The common reason offered as an explanation was that although CNTs have excellent intrinsic thermal conductivity they do not couple well to the matrix material or contact surface. The reported thermal boundary resistance (TBR) between CNTs and polymer matrix was as high as ~$10^{-7}$ m$^2$KW$^{-1}$.[12] The large TBR at the CNT – matrix interface can be attributed to the fundamental property – high Kapitza resistanceH[13] between one-dimensional (1D) CNTs and 3D bulk owing to the large difference in the phonon density of states (DOS). It was also suggested that the lack of thermal percolation in CNT composites can negatively affect their heat conduction properties.[7]

The effective utilization of CNTs in composites also depends on their ability to disperse individually and homogeneously within a matrix material. In order to maximize the effects of CNTs for improving the electrical, thermal and mechanical properties of polymer composites one should ensure that CNTs do not form aggregates and, instead, disperse into a network with increased accessible interfacial surface area.[14] Chemical functionalization has been considered a possible route for achieving this goal.[15] Subjecting CNTs to chemical treatment can improve the coupling at the CNT – matrix interface leading to enhanced electrical or thermal conductivity of the nanocomposites. However, chemical functionalization that works well for improving the electrical properties of nanocomposites may not necessarily result in improving the thermal properties. Moreover, the effects produced on SW-CNTs can be drastically different from those on multi-wall carbon nanotubes (MW-CNTs). The length and diameter of CNT can also affect the outcome of functionalization.

Here we report the results of the experimental investigation of the effects produced by the carboxylic functionalization (-COOH groups) of CNTs on the thermal properties of the resulting carbon-nanotube – polymer nanocomposites. The carboxylic functionalization (-COOH groups) of CNTs allows one to substantially improve the electrical properties of NCs.[16] The situation with the thermal properties of NCs is more complicated and the net effect depends on the interplay of improved CNT – matrix coupling and degraded intrinsic thermal conductivity of functionalized CNTs. The obtained results shed light on the phonon heat conduction in CNTs embedded inside the polymer matrix and can help finding a trade off when simultaneous increase in the electrical and thermal conduction properties of NCs is required.

**RESULTS OF THE MEASUREMENTS**

For this study, a set of composite samples was prepared with two different matrix materials: (i) thermoset commercial epoxy resin, utilized in the automotive industry, and (ii) polydimethylsiloxane (PDMS), a silicone rubber, which is widely used for bio-applications and as thermal interface materials (TIMs). The list of samples with brief description is given in Table I. The prepared composites contained various carbon materials as fillers, including SW-CNTs and MW-CNTs with different diameters and lengths. A set of SW-CNTs and MW-CNTs were subjected to carboxylic (-COOH) surface functionalization before addition to the composites while others were added as grown (see Figure 1 (a-c)). The amount of carbon fillers in NCs was varied as 0, 1, and 3 wt %. The composites were prepared with twelve different fillers. The functionalized filler (filler #8) was added to PDMS matrix at 3 wt% using similar methods as with the Epilox epoxy resin samples. Figure 1 (b) shows CNT – polymer NCs prepared in the form of cylinders of 12-mm diameter and 1-2-mm thickness for the thermal study.

[Table I]

An ideal SW-CNT possesses a continuous lattice of $sp^2$ bonded carbon. It has very high intrinsic thermal conductivity of ~3000 W/mK at room temperature (RT).[17] Owing to the extra strong $sp^2$ bonding, acoustic phonons are the main carrier of heat in all types of CNTs. The only other material system that revealed a higher thermal conductivity is graphene.[3,18-19] The carboxylic surface functionalization is carried out by placing CNTs in an oxidizing environment, *e.g.* nitric acid at ~130 ºC, for an extended period of time.[20] The rate and degree of carboxylic oxidation functionalization depends on the number of defect sites already present on CNTs. MW-CNTs are expected to have fewer defects than SWCNTs due to their inherently larger radius of curvature and correspondingly lower in-plane strain. It was observed that the oxidation of MW-CNTs occurs more slowly and selectively than that of SW-CNTs. In case of MW-CNTs, the oxidation occurs mainly at the nanotubes ends. As a result, SW-CNTs are expected to have more sidewall functionalization sites than MW-CNTs.[16,21] The sidewall functionalization and other contamination can adversely affect the electronic properties of CNTs.[22] This effect is stronger for SW-CNTs as all the current is carried by one wall. MW-CNTs are less sensible to this effect as the current can redistribute among other walls. The carboxylic functionalization is expected to improve filler dispersion within the bulk polymer matrix by reducing the hydrophobicity of CNT filler material.[20] For the same reason, the carboxylic functionalization is also expected to improve coupling between the CNT and surrounding polymer matrix. The carboxylic functionalization of the sidewalls of CNTs results in breaking of $sp^2$ bonds and can lead to formation of $sp^3$ covalent bonds or C-H defect sites that act as scattering centers for acoustic phonon propagating along CNTs (see Figure 1 (a)).

[Figure 1 (a-d)]

The previous studies of electrical properties of the functionalized CNT – polymer NCs, prepared by the same method, established that the electrical characteristics were isotropic suggesting that the filler is dispersed without preferential orientation.[14-15] This was directly confirmed by electron microscopy (Figure 1 (c)). It is also known that CNT NCs with functionalization reveal an increase in the electrical conductivity from essentially non-conductive ($\sim 10^{-11}$ S/cm) to $6 \times 10^{-7}$ S/cm at 3 wt% SW-CNT (filler #6) and $10^{-5}$ S/cm and $2 \times 10^{-2}$ S/cm at 1 and 3 wt% (filler #3), respectively. The electrical conductivity of MW-CNT NCs is increased by a factor of 7 at 1 wt% filler and by an additional $\sim 20\%$ at 3 wt% filler when carboxylic functionalization is applied (relative to MW-CNT NCs with unfunctionalized CNTs of the same type.[20,23] The electrical conductivity of the SW-CNT-composites was increased by 34% at 1 wt% filler but decreased by a factor of $\sim 3$ at 3 wt% filler when carboxylic functionalization is applied (relative to SW-CNT-composites with unfunctionalized CNTs of the same type[23] ).

The measurements of the cross-plane thermal conductivity of CNT – polymer NCs were performed using the laser flash technique (LFT). The details of the measurements procedures are provided in the Methods section. Figure 2 shows the thermal conductivity as a function of temperature for the neat epoxy, nanocomposite (NC) with SW-CNT (filler #6), NC with functionalized SW-CNT (filler #7), NC with MW-CNT (filler #3) and NC with functionalized MW-CNT (filler #8). The filler fraction in all NCs is 3 wt%. One can see that addition of as grown SW-CNTs to the epoxy enhances the thermal conductivity by about 25%.

Functionalization degrades the performance of SW-CNT fillers. The thermal conductivity drops almost to the neat epoxy value. This trend suggests that while increasing the SW-CNT coupling to the matrix the side-wall defect sites created by the functionalization strongly impede the acoustic phonon transport along the SW-CNT. Addition of the same fraction of MW-CNTs also enhances the NC thermal conductivity although not as strongly as SW-CNTs. Functionalization of MW-CNTs does not produce a noticeable effect on the enhancement factor. The latter can indicate that the contribution of the outer wall to the heat conduction in MW-CNTs is marginal and heat transport goes *via* the inside walls, which are not affected by the defects created by functionalization on the outer wall. It is illustrative to compare the functionalization effect on thermal and electrical properties[20]. At 3 wt% loading of SW-CNT (filler #7) the functionalization reduced the thermal conductivity and decreased the electrical conductivity. At 3 wt% loading of MW-CNT (filler #8), the functionalization had little effect on the thermal conductivity while increasing the electrical conductivity by an additional ~20% compared to the unfunctionalized MW-CNT of the same type.

[Figure 2]

Figure 3 presents the thermal conductivity for the neat epoxy, neat PDMS and two types of NCs. The data are shown for 1 wt% and 3 wt% of the epoxy – MW-CNT NC (filler #16) and 3 wt% of the PDMS – functionalized MW-CNT NC (filler #8). The results on this plot follow the expected trend. Addition of a larger fraction of CNTs enhances the thermal conductivity stronger. The PDMS-based NCs loaded with CNTs reveal higher *K*. The data presented in Figure 3 allows one to analyze the temperature dependence of the thermal conductivity. The epoxy-based NCs reveal a non-monotonic K *vs.* T dependence, which likely reflects the heat capacity behavior. The specific heat of epoxy can change with temperature owing to the loss of the water adsorbed from the ambient or additional curing during heating. The epoxy tends to absorb water stronger than PDMS saturating at 1.5-6 wt%.[24] The thickness of the samples is sufficiently small to allow for humidity effect on the epoxy during the sample storage. The effect of adsorbed water on properties varies with the chemistry of the epoxy resin and hardener. However, the loss of water during the thermal measurements at different elevated temperature would lower the specific heat (the specific heat of water is 4 J/g·K while that of the epoxy/air is 1 J/g·K) resulting in a valley around 65 – 75°C. PDMS has a more monotonically increasing specific heat with the temperature in comparison with that of the epoxy.

[Figure 3]

Figure 4 (a-b) shows the effect of the length of the as-grown CNT on the thermal conductivity. The data in Figure 4 (a) is presented for the SW-CNTs with the 3 wt% loading fraction. The "long" SW-CNTs (filler #4) enhance K stronger than the "short" SW-CNTs (filler #5). The same trend is observed for the MW-CNTs presented in Figure 4 (b). The difference in the length of CNT in both cases is on the order of one order of magnitude. The estimated average ratio of the CNT length, $L$, to diameter, $D$, in the case of SW-CNTs was $L/D \sim 10^4$ for the "long" SW-CNTs and $L/D \sim 10^3$ for the "short" SW-CNTs. In the case of the "long" MW-CNTs, the ratio $L/D$ was above $\sim 10^3$ while for the "short" MW-CNTs, the ratio was $\sim 10^2$. One can conclude that longer CNTs with larger $L/D$ are more efficient in improving the heat conduction properties of CNT – polymer NCs. This conclusion is supported with the calculations reported for composites with MW-CNTs using the Hatta-Taya model for randomly distributed fibers.[25]

[Figure 4 (a-b)]

In Figure 5, we present the thermal conductivity enhancement as a function of diameter, $D$, at two different MW-CNT loading fractions. Despite some data scatter the data suggests a trend that larger $D$ is beneficial for the thermal conductivity of MW-CNT-polymer NCs. In previous studies of the electrical conductivity of CNT composites it was found that the enhancement effect was volumetric.[25]   In this case, the MW-CNTs all have similar volume fractions ~0.6 and 1.7 vol % at 1 and 3 wt% respectively. It was also found that the enhancement trend depended more on diameter of the filler rather than *L/D* ratio.

[Figure 5]

## DISCUSSION

For practical applications, it is important to identify the CNT fillers and loading fractions, which improve the electrical conductivity without degrading, or preferably, simultaneously increasing the thermal conductivity. Figure 6 presents the electrical conductivity data[20] and our measured thermal conductivity enhancement factors (see Eq.(1)) for the epoxy-based NCs with the as-grown SW-CNTs (filler #6; red line), functionalized SW-CNTs (filler #7; blue line), as-grown MW-CNTs (filler #3; brown line) and functionalized MW-CNTs (filler #8; green line). One should note that addition of 1 wt% of CNTs is usually not sufficient to improve the thermal conductivity of NCs and can even result in its degradation. Two data points with negative $\eta$ are shown in Figure 6. Four of the 11 tested fillers (samples #4, #5, #7, and #8) revealed a decrease in the average thermal conductivity at 1 wt% compared to the neat epoxy over the examined temperature range. It is interesting to note that in the sample #8 addition of 1 wt% of MW-CNTs improved the electrical properties of the composite substantially while slightly decreasing the thermal conductivity from the neat epoxy value. One can conclude from this data that in order to ensure preservation or enhancement of the thermal conduction properties of CNT – polymer NCs the filler loading should be above 1 wt%, preferably closer to 3 wt%.

One can see from Figure 6, that in case of SW-CNTs, functionalization produces negative effect on the thermal conductivity. The side-wall defects created by functionalization improve the SW-CNT coupling to the matrix but deteriorate the thermal conductivity of SW-CNT itself. The defects act as scattering centers for the acoustic phonons that carry the bulk of heat in all CNTs.[3,18,26] In case of MW-CNTs, functionalization improves dispersion and electrical properties without degrading the thermal conductivity, which is primarily determined by the filler loading fraction. Note that the sample #8 with the functionalized MW-CNT filler at 3 wt% loading demonstrated the highest electrical conductivity of 25 mS/cm with the thermal conductivity simultaneously enhanced by $\eta \sim 15\%$. The latter can be explained assuming that all walls contribute to the phonon transport so that the increase in defectiveness of the external wall consequent to functionalization leads to a marginal decrease in thermal conductivity overwhelmed by the decrease in thermal contact resistance due to functionalization. Thus, -COOH functionalized MW-CNTs appear to be the optimum filler for the polymer NCs when improved uniformity of the filler dispersion is desired with the simultaneous enhancement of the electrical and thermal conduction properties.

In this study, we specifically limited our analysis to relatively low loading fractions of CNTs. From the technological point of view, we were motivated to find a practically feasible way of increasing both the electrical and thermal properties. The larger CNT loading fractions are impractical because of the prohibitive cost of CNTs and additional cares needed to achieve proper dispersion. Higher loading usually results in less uniform CNT dispersion making comparison less accurate. It was also observed earlier[20] that the electrical conductance in our samples starts to saturate for the loading fractions above 3 wt%.

Although a direct comparison of the results is not possible due to differences in the composition and preparation methods, it is illustrative to discuss our data in the context of prior studies of the thermal and electrical properties of CNT – polymer composites.[21,27-29] It was reported that subjecting MWCNTs to triethylenetetramine (TETA) grafting was beneficial for the epoxy-based composites at all investigated loading fractions.[21] However, TETA functionalization is different from the one used in our study. The authors indicated that TETA-functionalized MWCNTs constitute the "core-shell" structure (MWCNT is the core and TETA is the shell), which explain its benefit for enhancing the thermal conduction and mechanical strength.[21] Our task was to test if increasing the CNT – matrix coupling leads to simultaneous enhancement of the electrical and thermal properties. Another study reported increased thermal conductivity in the silicone composite with 2 vol% of MWCNTs.[30] The examined silicone matrix is similar to our PDMS while the CNT length is much shorter leading to much lower CNT distances. Our results for thermal properties of CNT – PDMS composites are in line with the data reported in Ref. [30]. The thermal conductivity measurements of purified SW-CNT – epoxy composites prepared using suspensions of SW-CNTs in N-N-Dimethylformamide (DMF) and surfactant stabilized aqueous SW-CNT suspensions revealed some enhancement of the thermal conductivity at rather small weight fractions (~1%) of SW-CNTs.[27] However, the SW-CNTs in this work were not subjected to chemical functionalization and had much smaller length than CNTs in our study. Finally, based on our results and data reported in literature[7,28-29] one can observe that the question of the thermal percolation remains open.[7,28-29]

**CONCLUSIONS**

We have studied experimentally how the carboxylic surface functionalization of CNTs affects the thermal conductivity of the CNT – polymer NCs. It was found that the thermal conductivity of polymers with the functionalized carbon nanotubes is weakly or positively affected in the case of the multi-wall carbon nanotubes, while it is reduced in the case of the single-wall carbon nanotubes. Improvement of thermal conduction requires larger loading of CNTs than those needed for increasing the electrical conductivity. The longer and larger diameter carbon nanotubes are more efficient for enhancing the thermal conductivity of polymer composites. The observed trends were explained by the fact that while surface functionalization increases the coupling between carbon nanotube and polymer matrix it also leads to formation of defects, which impede the acoustic phonon transport in the single wall carbon nanotubes.

**METHODS**

The composites for this study were prepared with several different CNT fillers. The desired quantity of filler was mixed into liquid epoxy resin stirred vigorously (1200-rpm for 10 min.)

The sonication with the ultrasonic frequency 37 KHz for 15 min was applied to help remove voids from the mixture. Degassing in vacuum was performed to completely remove air pockets. The composite mixture was poured into a mold and cured in oven at 70°C for 4 hours. A more detailed description of the samples was reported by some of us elsewhere.[20,23,31]

In order to perform the measurement of the thermal conductivity with LFT, each sample was placed into a stage that fitted its size. The bottom of the stage was illuminated by the flash of a xenon lamp. The temperature of the top surface of the sample was monitored by a cryogenically cooled InSb IR detector. The temperature rise as a function of time allows the determination of the thermal diffusivity. The thermal conductivity, $K$ (W/m K), is calculated from the equation

$$K = \alpha \, \rho \, c_p \qquad (2)$$

where $\alpha$ is the thermal diffusivity (m/s$^2$), $\rho$ is the mass density of the material (kg/m³) and $c_p$ the specific heat (J kg$^{-1}$ K$^{-1}$). The specific heat was obtained from the laser flash calorimetry.[32] The method requires a reference sample of known heat capacity and similar thermo-physical properties. In our case the neat epoxy and neat PDMS were used as the references. An assumption was made that the reference sample and the test sample absorb an equal amount of energy from equal light flashes. Further details of the laser flash measurement procedures, in the context of other material systems, have been reported elsewhere.[33-34]

## Acknowledgements

The work in Balandin Group at UCR was supported by the Semiconductor Research Corporation (SRC) and the Defense Advanced Research Project Agency (DARPA) through FCRP Center for Function Accelerated nanoMaterial Engineering (FAME). The work in the Carbon Group at Polytechnic of Turin has been supported by the THERMALSKIN Project on "Revolutionary surface coatings by carbon nanotubes for high heat transfer efficiency". The authors thank Dr. S. Guastella at Polytechnic of Turin for his FESEM measurements. The authors also acknowledge Nanothinx for providing their CNTs free of charge.

**References**

[1] Breuer, O.; Sundararaj, U. Big Returns from Small Fibers: A Review of Polymer/Carbon Nanotube Composites. *Polym. Compos.* **2004**, *25*, 630-645.

[2] Coleman, J.N.; Khan, U.; Blau, W. J.; Gun'ko, Y. K. Small But Strong: A Review of the Mechanical Properties of Carbon Nanotube Polymer Composites. *Carbon* **2006**, *44*, 1624-1652.

[3] Balandin, A. A. Thermal Properties of Graphene and Nanostructured Carbon Materials. *Nat. Mater.* **2011**, *10*, 569 − 581.

[4] Shahil, K.M.F. and Balandin, A.A. Thermal Properties of Graphene and Multilayer Graphene: Applications in Thermal Interface Materials. *Solid State Commun.* **2012**, *152* , 1331 − 1340 (2012).

[5] MacDiarmid, A. G. Synthetic Metals: A Novel Role for Organic Polymers. *Synth. Met.* **2002**, *125*, 11-22.

[6] Felba, J. Thermally Conductive Nanocomposites. In *Nano-Bio-Electronic, Photonic and MEMS Packaging*; Wong, C. P.; Moon, K.S.; Li, Y., Eds.; Springer Science: New York, 2010; pp. 277−314.

[7] Shenogina, N.; Shenogin, S.; Xue, L.; Keblinski, P. On the Lack of Thermal Percolation in Carbon Nanotube Composites. *Appl. Phys. Lett.* **2005**, *87*, 133106.

[8] Yu, A.; Itkis, M. E.; Bekyarova, E.; Haddon, R. C. Effect of Single-Walled Carbon Nanotube Purity on the Thermal Conductivity of Carbon Nanotube-Based Composites. *Appl. Phys. Lett.* **2006**, *89*, 133102.

[9] Bonnet, P.; Sireude, D.; Garnier, B.; and Chauvet, O. Thermal Properties and Percolation in Carbon Nanotube-Polymer Composites. *Appl. Phys. Lett.* **2007**, *91*, 201910.

[10] Choi, S. U. S.; Zhang, Z. G.; Yu, W.; Lockwood, F. E.; Grulke, E. A. Anomalous Thermal Conductivity Enhancement in Nanotube Suspensions. *Appl. Phys. Lett.* **2009**, *79*, 2252-2254.

[11] Moisala, A.; Lia, Q.; Kinlocha, I. A.; Windle, A. H. Thermal and Electrical Conductivity of Single and Multi-walled Carbon Nanotube Epoxy Composites. *Compos. Sci. Technol.* **2006**, *66*, 1285-1288.

[12] Huxtable, S.; Cahill, D. G.; Shenogin, S.; Xue, L.; Ozisik, R.; Barone, P.; Usrey, M.; Strano, M. S.; Siddons, G.; Shim, M.; *et al*. Interfacial Heat Flow in Carbon Nanotube Suspensions. *Nat. Mater*. **2003**, *2*, 731-734.

[13] Kapitza, P. L. The Study of Heat Transfer in Helium II. *J. Phys. USSR*. **1941**, *4*, 181.

[14] Ostojic, G. N.; Hersam, M.C. Biomolecule-Directed Assembly of Self-Supported, Nanoporous, Conductive, and Luminescent Single-Walled Carbon Nanotube Scaffolds. *Small*. **2012**, *8*, 1840-1845.

[15] Ma, P.C.; Siddiqui , N. A.; Marom, G.; Kim, J.K. Dispersion and Functionalization of Carbon Nanotubes for Polymer-based Nanocomposites: A Review. *Composites: Part A* **2010**, *41*, 1345–1367.

[16] Kwon, J.; Kim, H. Preparation and Properties of Acid-treated Multiwalled Carbon Nanotube/Waterborne Polyurethane Nanocomposites. *J. Appl. Polym. Sci.* **2005**, *96*, 595-604.

[17] Kim, P.; Shi, L.; Majumdar, A.; McEuen, P. L. Thermal Transport Measurement of Individual Multiwalled Nanotubes. *Phys. Rev. Lett*. **2001**, *87*, 215502-215504.

[18] Nika, D.L.; Balandin, A. A. Two-Dimensional Phonon Transport in Graphene. *J. Phys.: Condens. Matter.* **2012**, *24*, 233203.

[19] Balandin, A. A.; Nika, D.L. Phonons in Low-dimensions: Engineering Phonons in Nanostructures and Graphene. *Materials Today*. **2012**, http://www.materialstoday.com.

[20] Chiolerio, A.; Castellino, M.; Jagdale, P.; Giorcelli, M.; Bianco, S.; Tagliaferro, A. Electrical Properties of CNT-Based Polymer Matrix Nanocomposites. In *Carbon Nanotubes-Polymer Nanocomposites*; Yellampalli, S., Ed.; InTech: 2011; pp 215-230.

[21] Yang, K.; Gu, M.; Yuping, G.; Pan, X.; Mu, G. Effects of Carbon Nanotube Functionalization on the Mechanical and Thermal Properties of Epoxy Composites. *Carbon*, **2009**, *47*, 1723-1737.

[22] Berger, C.; Yi, Y.; Wang, Z. L.; de Heer, W. A. Multiwalled Carbon Nanotubes are Ballistic Conductors at Room Temperature. *Appl. Phys. A: Mater. Sci. Process.* **2002**, *74*, 363-365.

[23] Castellino, M.; Chiolerio, A; Shahzad, M.I.; Jagdale, P.; Tagliaferro, A. Electrical

Conductivity Phenomena in an Epoxy Resin-Carbon-Based Materials Composite. Submitted.

[24] Voo, R.; Mariatti, M.; Sim, L.C. Thermal Properties and Moisture Absorption of Nanofillers-Filled Epoxy Composite Thin Film for Electronic Application. *Polym. Adv. Technol*. **2012**, *23*, 1620-1627.

[25] Gojny, F. H.; Wichmann, M. H. G.; Fiedler, B.; Kinloch, I. A.; Bauhofer, W.; Windle, A. H.; Schulte, K. Evaluation and Identification of Electrical and Thermal Conduction Mechanisms in Carbon Nanotube/Epoxy Composites. *Polymer*. **2006**, *47*, 2036-2045.

[26] Lee, J.; Varshney, V.; Roy, A. K.; Farmer, B.L. Single Mode Phonon Energy Transmission in Functionalized Carbon Nanotubes. *J. Chem. Phys.* **2011**, *135*, 104109.

[27] Bryning, M.B.; Milkie, D.E.; Islam, M.F.; Kikkawa, J.M.; Yodh, A.G. Thermal Conductivity and Interfacial Resistance in Single-Wall Carbon Nanotube Epoxy Composites. *Appl. Phys. Lett.* **2005**, 87, 161909.

[28] Bryning, M.B.; Islam, M.F.; Kikkawa, J.M.; Yodh, A.G. Very Low Conductivity Threshold in Bulk Isotropic Single-Walled Carbon Nanotube-Epoxy Composites. *Adv. Mater.* **2005**, 17, 1186-1191.

[29] Kilbride, B.E.; Coleman, J.N.; Fraysse, J.; Fournet, P.; Cadek, M.; Drury, A; Hutzler, S.; Roth, S.; Blau, W.J. Experimental Observation of Scaling Laws for Alternating Current and Direct Current Conductivity in Polymer-Carbon Nanotube Composite Thin films. *J. Appl. Phys.* **2002**, 92, 4024-4030.

[30] Lee, G.; Lee, J. I.; Lee, M. P.; Kim, J. Comparisons of Thermal Properties between Inorganic Filler and Acid-Treated Multiwall Nanotube/Polymer Composites. *J. Mat. Sci.* **2005,** 40, 1259-1263.

[31] Shahzad, M. I. ; Shahzad, N.; Giorcelli, M.; Tagliaferro, A. Optical Performance of Carbon Nanotubes based Polydimethylsiloxane Composite Films. Submitted

[32] Gobel, A.; Hemberger, F.; Vidi, S.; Ebert, H.P. A New Method for the Determination of the Specific Heat Capacity Using Laser-Flash Calorimetry Down to 77K. *Int. J. Thermophys.* **2012**.

[33] Ghosh, S.; Teweldebrhan, D.; Morales, J. R.; Garay, J. E.; Balandin, A. A. Thermal Properties of the Optically Transparent Pore-Free Nanostructured Yttria-stabilized Zirconia. *J.*

*Appl. Phys*. **2009**, *106*, 113507.

[34] Yan, Z.; Liu, G.; Khan, J.M.; Balandin, A.A. Graphene Quilts for Thermal Management of High-power GaN Transistors. *Nat. Commun.* **2012**, *3*, 827.

**FIGURE CAPTIONS**

**Figure 1:** (a) Illustration of carboxylic groups attached to the side-wall of SW-CNT. The new bond interrupts the continuous $sp^2$ crystal lattice of SW-CNT. The carboxylic groups usually form at the sites of the dangling bonds, defects or CNT ends. While increasing the CNT – matrix coupling, the attachment of functionalized groups creates scattering centers for acoustic phonons carrying heat in CNTs. (b) Photo of CNT – polymer composites prepared in the form of cylinders for thermal measurements. (c) Scanning electron microscopy image of epoxy-based nanocomposite with 3 wt-% of MW-CNTs (sample #14) showing the random orientation of CNTs. The scale bar in the image is 300 nm. (d) Scanning electron microscopy image of epoxy-based nanocomposite with 3 wt-% of MW-CNTs (sample #16). The scale bar in the image is 1 μm.

**Figure 2:** Thermal conductivity of the epoxy-based nanocomposites with functionalized and unfunctionalized SW-CNT and MW-CNT (fillers # 3, 6, 7 and 8) at 3 wt% loading fraction. Note that functionalization produces little effect on the thermal conductivity in case of MW-CNT filers. The thermal conductivity enhancement is degraded by functionalization of SW-CNT fillers.

**Figure 3:** Thermal conductivity of PDMS with and without filler #8 at 3 wt% as a function of temperature. The thermal conductivity of epoxy composites with 1 and 3 wt% filler #16 is also plotted for comparison.

**Figure 4:** (a) Thermal conductivity of epoxy-based nanocomposites with SW-CNTs of different length. (b) The thermal conductivity of epoxy-based nanocomposites with MW-CNTs of different length. Note that longer CNTs perform better in terms of the thermal conductivity enhancement in both cases.

**Figure 5:** Thermal conductivity enhancement factor as a function of the outer diameter of MW-CNT filler at two different loading fractions. Each pair of points with the same diameter on the graph corresponds to the same filler type. In order of increasing diameter the data points correspond to the fillers #2, #16, #14, #15 and #1, respectively.

**Figure 6:** Electrical conductivity *vs.* the thermal conductivity enhancement factor for the epoxy based nanocomposites with the functionalized and unfunctionalized SW-CNT and MW-CNT fillers at 1 and 3 wt% loading fraction. The electrical conductivity is shown in the logarithmic scale. The data points correspond to the fillers # 3, 6, 7 and 8.

**Table I: Characteristics of the Carbon Nanotube Samples**

| Sample # | Filler type | Maker | Diameter (nm) | Length (μm) | Purity (weight %) |
|---|---|---|---|---|---|
| 1 | Multi Wall | Cheaptubes | 30-50 | 10-20 | > 95 |
| 2 | Multi Wall | Cheaptubes | < 8 | 10-30 | > 95 |
| 3 | Short Thin MW | Nanocyl 3150 | 9.5 | 1.5 | > 95 |
| 4 | Single Wall | Cheaptubes | 1-2 | 5-30 | > 90 |
| 5 | Single Wall | Cheaptubes | 1-2 | 0.5-2 | > 90 |
| 6 | Single Wall | Nanocyl 1100 | 2 | several | > 70 |
| 7 | -COOH Functional SW | Nanocyl 1101 | 2 | several | > 70 |
| 8 | -COOH Functional MW | Nanocyl 3100 | 9.5 | 1.5 | > 95 |
| 14 | Multi Wall | Nanothinx NTX-1 | 18-35 | >10 | 97 |
| 15 | Multi Wall | Nanothinx NTX-3 | 25-45 | >10 | 98.5 |
| 16 | Multi Wall | Nanothinx NTX-4 | 6-10 | >10 | >90 |

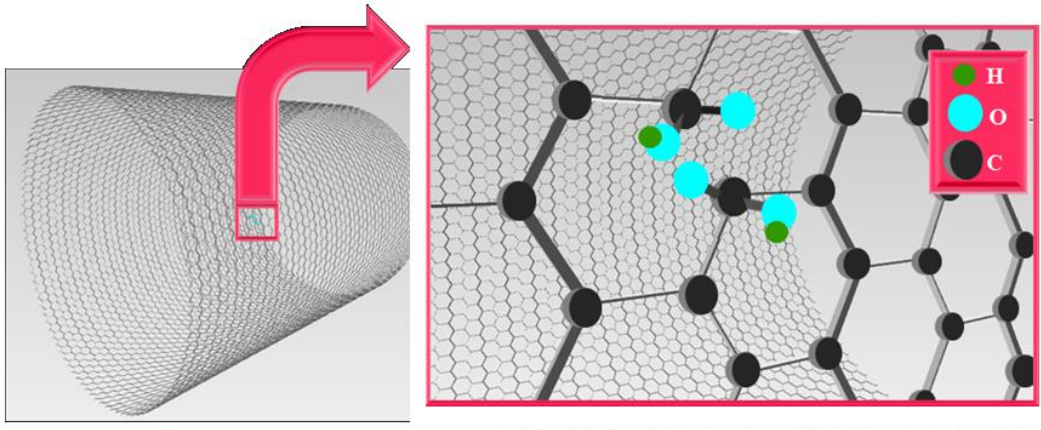

(a)

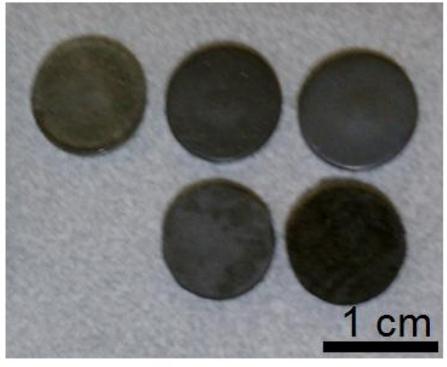

(b)

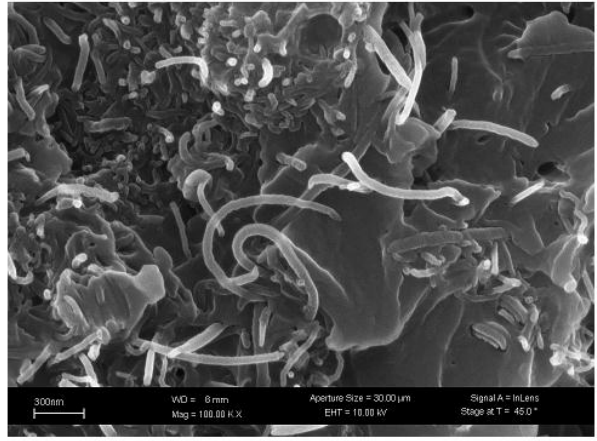

(c)

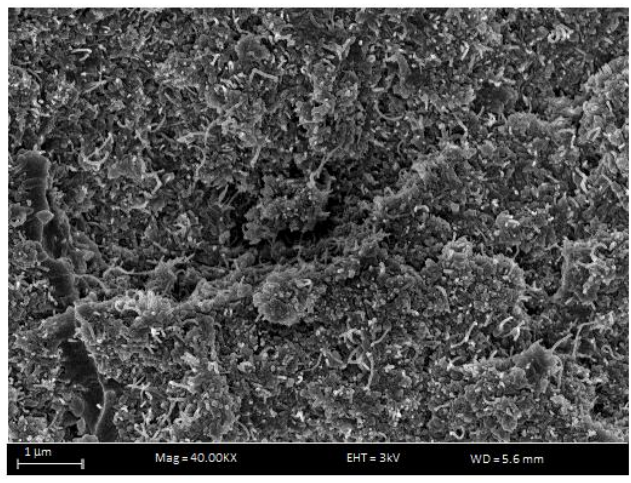

(d)

Figure 1

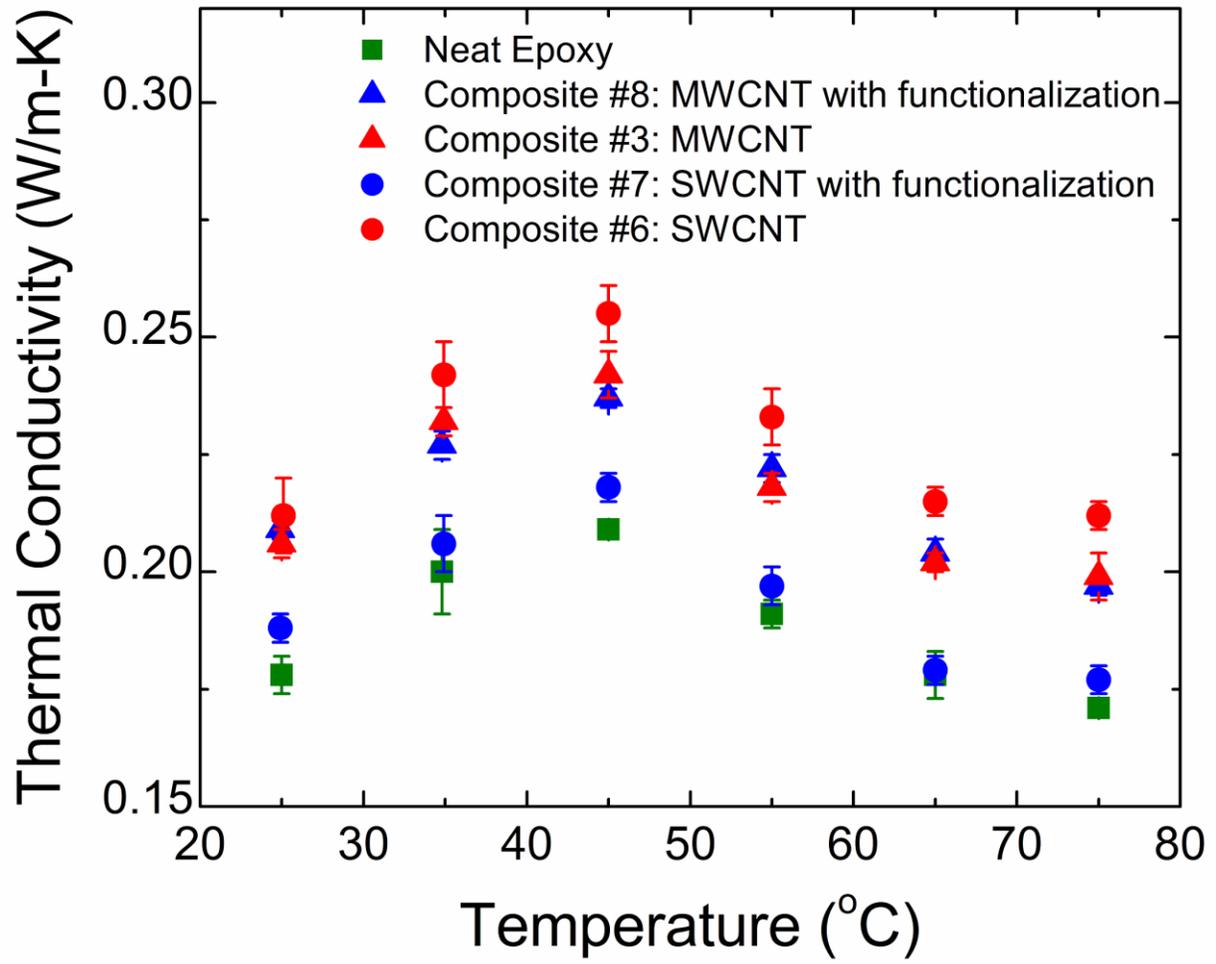

Figure 2

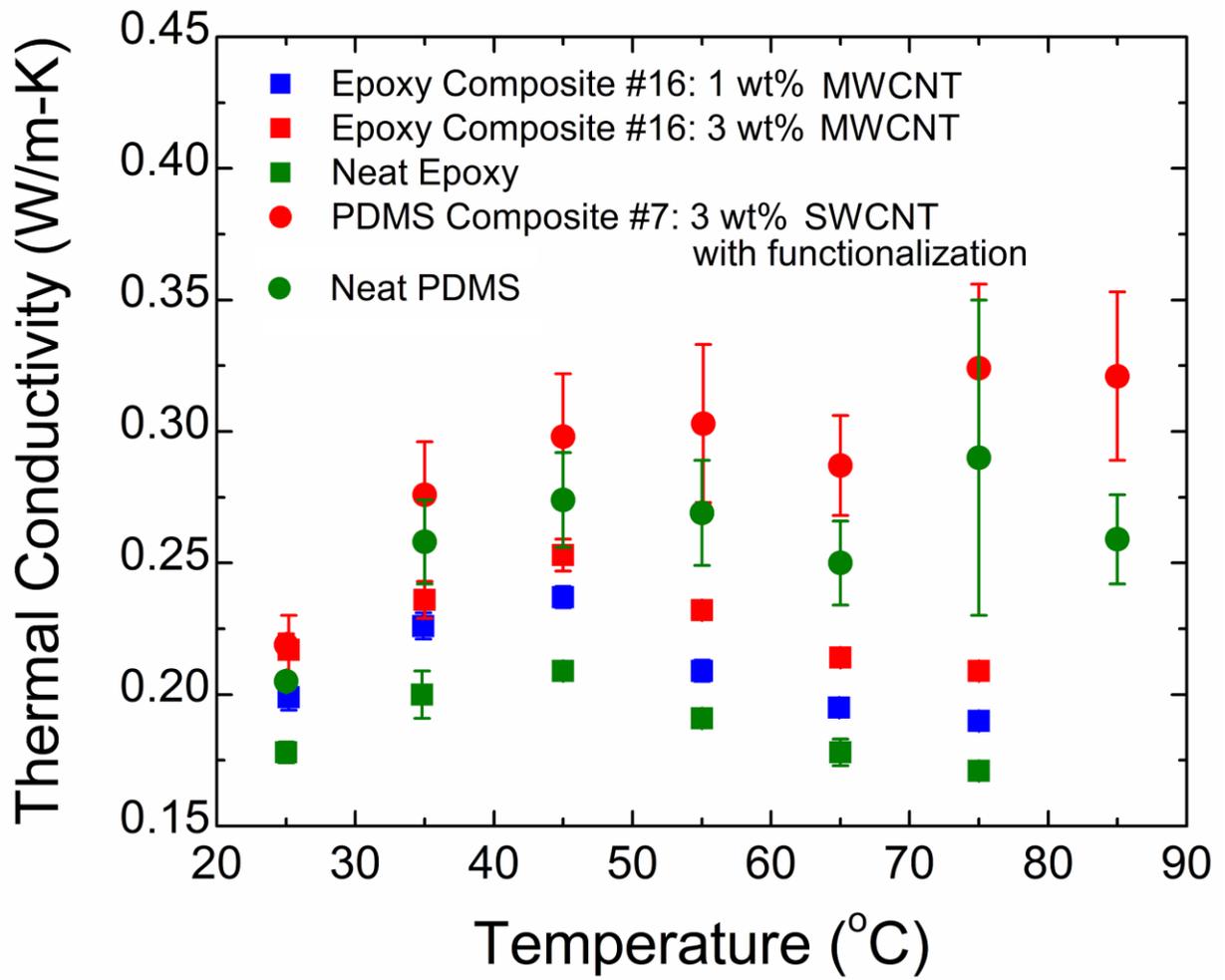

Figure 3

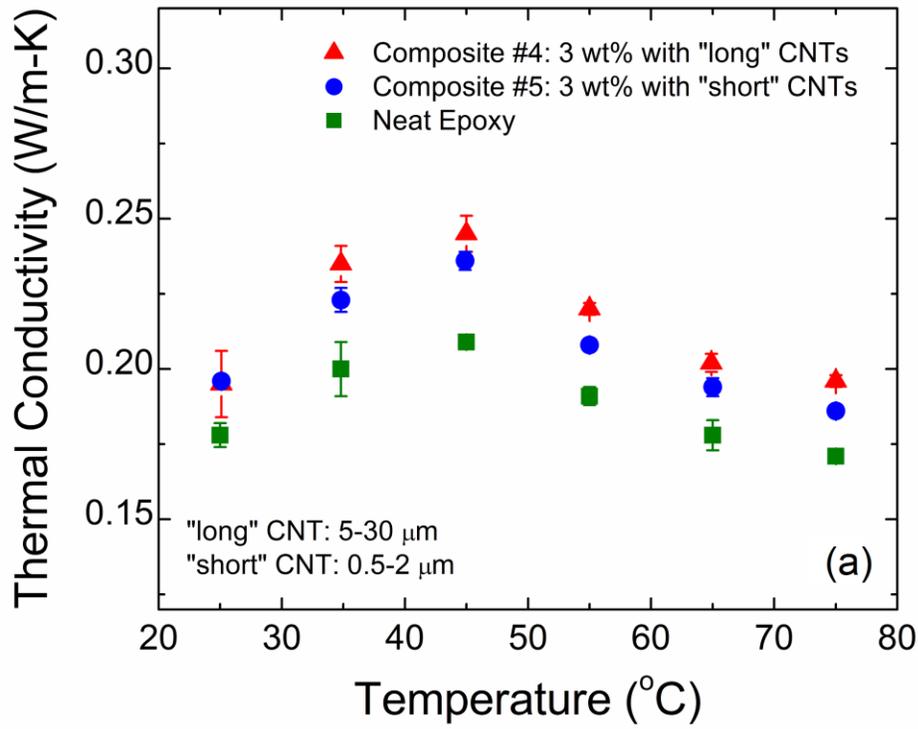

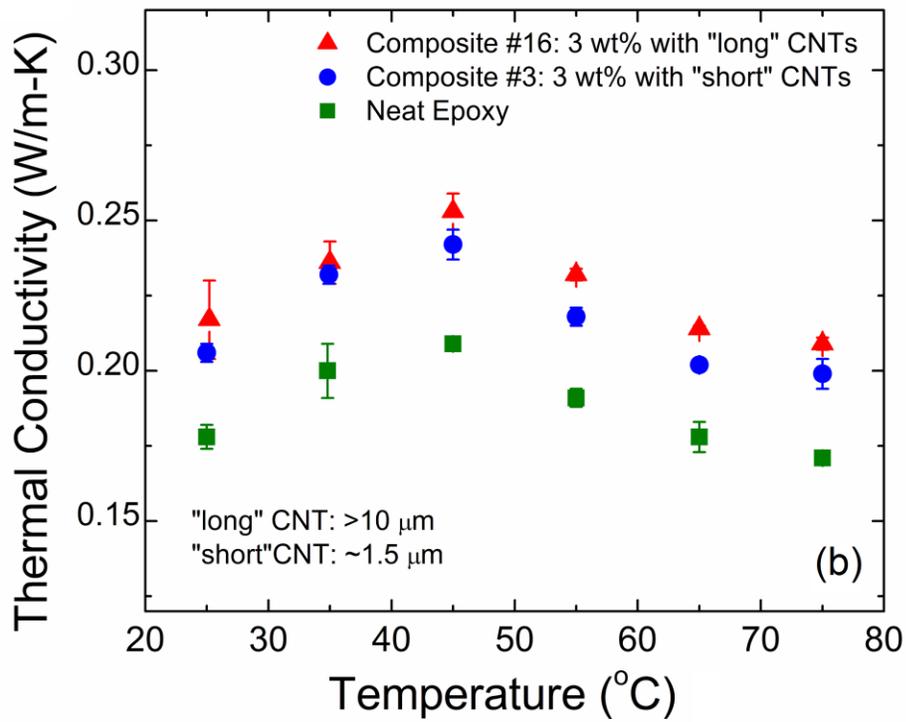

Figure 4

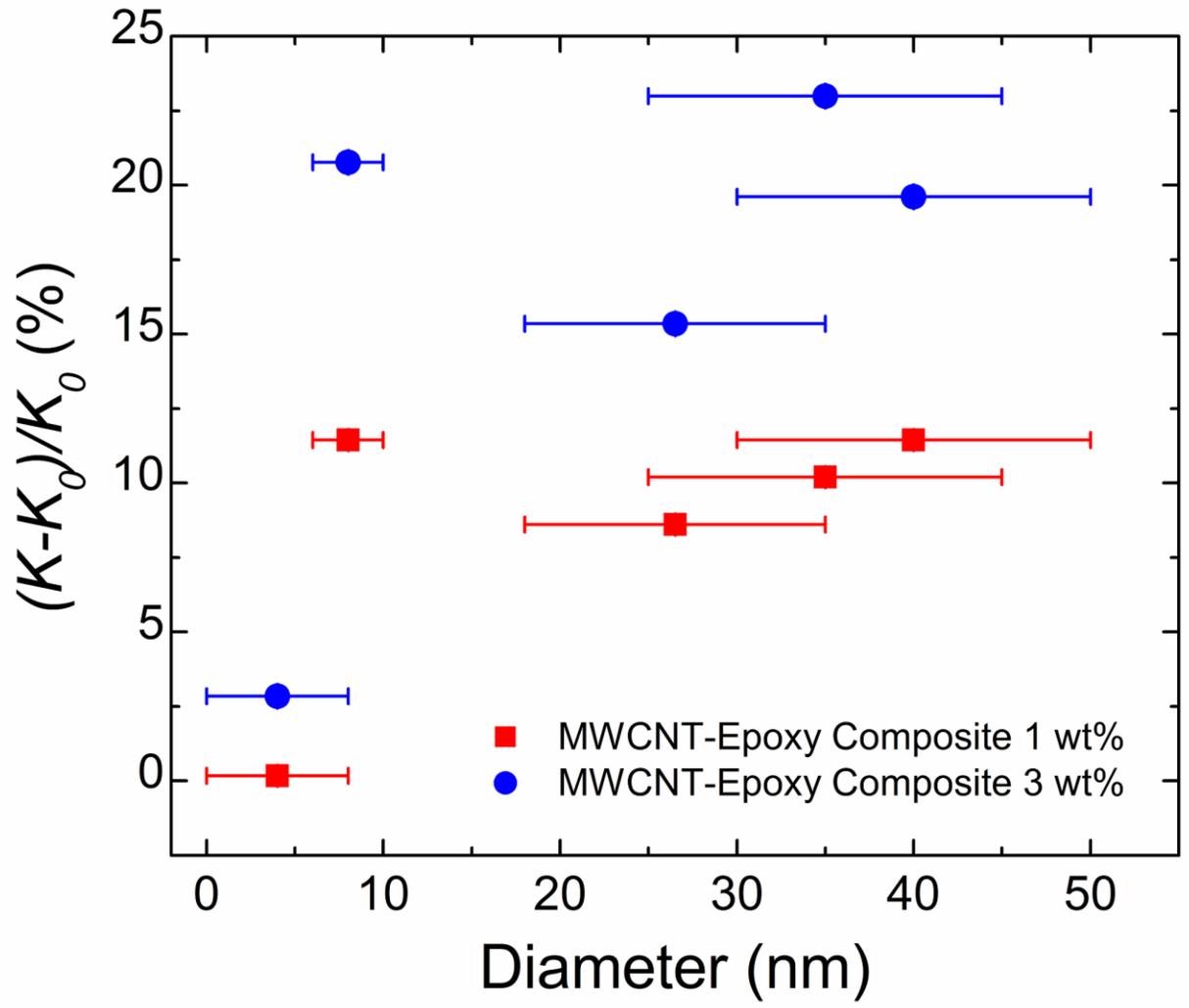

Figure 5

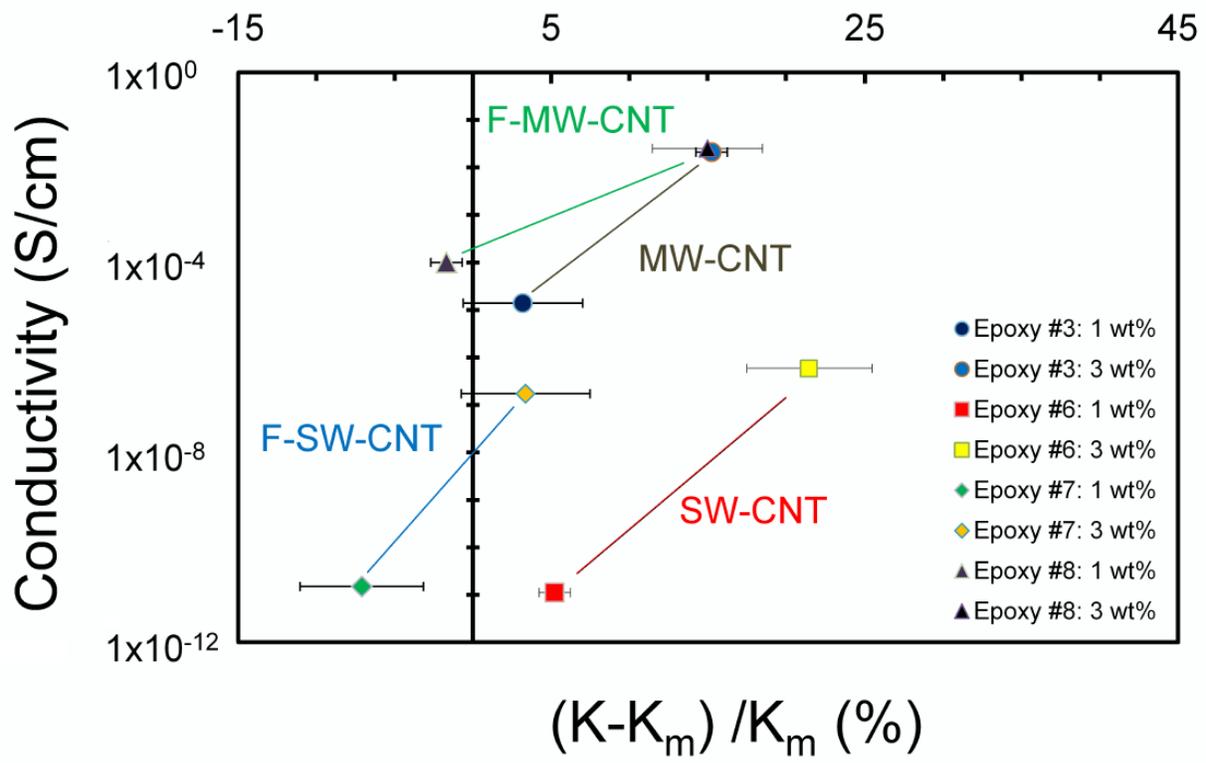

Figure 6